\documentclass{elsart3p}
\usepackage{natbib}
\usepackage{epsfig}
\usepackage{subfigure}
\usepackage{amsmath}
\usepackage{amssymb}

\def\mnras{MNRAS}
\def\apj{ApJ}

\def\aap{A\&A}

\def\prd{Phys.~Rev.~D}

\begin{document}

\begin{frontmatter}



\title{Effects of primordial helicity on CMB}


\author[label1,label2]{T. Kahniashvili}

\address[label1]{Department of Physics, Kansas State University, 116 
Cardwell Hall, Manhattan, KS 66502, USA}
\address[label2]{Center for Plasma Astrophysics, Abastumani Astrophysical Observatory, A. Kazbegi ave. 2a, Tbilisi, Ge-0160, Georgia}
\ead{tinatin@phys.ksu.edu}

\date{May 2006 \hspace{0.3truecm} KSUPT-06/3}
\begin{abstract}
I present here a brief 
  overview of the effects caused by parity violating 
cosmological sources (such as magnetic or kinetic helicity) on the 
CMB fluctuations. I discuss also primordial helicity induced 
relic gravitational waves. All these effects can serve as 
cosmological tests for 
primordial helicity detection. 
\end{abstract}


\end{frontmatter}

\section{Introduction}
\label{introduction}
Current and future  measurements of cosmic 
microwave background (CMB) temperature and polarization anisotropies 
(Page et al. 2006, Spergel et al. 2006)  
 provides probes physical processes in the early universe 
and 
 cosmological models. 
There are several astrophysical observations that indicate the presence of 
an helical magnetic field in clusters of galaxies (Widrow 2002,
Vall{\'e}e 2004, Semikoz and Sokoloff 2005a). A 
promising possibility to explain such a magnetic field is to  
assume  primordial 
 helicity generated during an early epoch of the universe expansion
 (Cornwall 1997, Giovannini \& Shaposhnikov 1998, Giovannini 2000,
Field \& Carroll 2000, Vachaspati 2001, Sigl 2002, Semikoz \&
Sokoloff 2005b, Campanelli \& Giannotti 2005). 
Conventionally we can distinguish two different kinds of helicity, 
kinetic helicity related to primordial plasma motions 
and magnetic helicity related to a primordial magnetic field 
 (Brandenburg 2001, Christensson, Hindmarsh, and
Brandenburg 2005, Verma \& Ayyer 2003, Boldyrev and Cattaneo 2004, Subramanian 2004). 
 
The average energy density and  
helicity of the magnetic have to be small enough to preserve spatial large-scale
isotropy of the universe. Under such assumptions  the linear perturbation 
theory of gravitational instability may be used to describe
perturbation the dynamics (for a review see Giovannini 2006a). 
 Of course,  the two  kinds of helicity are related 
 through magnetohydrodynamical evolution. 
On the other hand,  primordial kinetic helicity influences 
the dynamics of cosmological perturbation  (Vishniac \& Cho 2001,
Brandenburg 2001, Kleorin et al. 2003, Subramanian 2002, Vishniac,
Lazarian, and Cho 2003, Subramanian \& Brandenburg 2004, Banerjee
\& Jedamzik 2004, Subramanian, Shukurov and Haugen 2005), therefore
 it should be accounted
for when the cosmological effects of a primordial helical 
 magnetic field or/and 
of primordial helical turbulent motions 
are studied. 

The energy-momentum tensor associated with a primordial helicity source  
(e. g., a magnetic field or turbulent motions) 
 induces all modes of perturbations
(scalar, vector, and tensor). Neglecting
 second order effects (Bartolo, Mattarrese, and Riotto 2004, 
Lesgourgues et al, 2005) and the coupling between
scalar, vector, and tensor modes (which results in non-gaussianity of
the CMB fluctuations, Brown \& Crittenden,  2005), 
 the scalar, vector, and tensor modes can be studied separately. 

Here I focus on the effects on CMB temperature and polarization 
anisotropies induced by primordial helicity. This presentation is based on
results obtained in collaboration with C. Caprini,
R. Durrer, G. Gogoberidze, A. Kosowsky, G. Lavrelashvili, A. Mack,
and B. Ratra.  (see  Mack, Kahniashvili, and Kosowsky 2002,
Kosowsky, Mack and Kahniashvili 2002, 
Caprini, Durrer, and Kahniashvili 2004, Kosowsky et al. 2005,
Kahniashvili and Ratra 2005, Kahniashvili, Gogoberidze, and Ratra 2005, 
Kahniashvili and Ratra 2006).
We find that primordial helical sources generate vector and tensor
metric perturbations (primordial helicity does not influence the scalar 
mode of perturbations, Kahniashvili and Ratra 2006, while the energy 
density of the corresponding source does) and as a result affect 
all   
CMB fluctuations. In particular, 

(a)
Parity violation in the universe results in an asymmetry in the amplitude 
of left- and right-handed gravitational waves 
(Lue, Wang and Kamionkowski 1999). As a result 
 primordial helicity generates  circularly polarized
stochastic gravitational waves 
(Caprini et al. 2004, Kahniashvili et al. 2005), 
which can be directly detected 
by future  space based gravitational wave detection missions.
Cosmological helicity also induces parity 
violating vorticity pertubations (Pogosian. Vachaspati, and Winitzki  2002, 
Kahniashvili \& Ratra 2005). 

  (b) Cosmological helicity 
reduces the amplitudes of  the parity-even CMB fluctuation
power spectra compared to the case 
without primordial helicity 
(Caprini et al. 2004,
Kahniashvili and Ratra 2005, Kahniashvili, et al. 2005);

(c) Faraday rotation of the 
CMB polarization plane is strongly dependent on the average 
energy density of 
the cosmological magnetic field and is independent of magnetic helicity 
(Kosowsky et al. 2005). The scalar mode of perturbations does not reflect 
the presence of primordial helicity (Kahniashvili \& Ratra 2006). These 
features of primordial helicity can be used as additional tests 
when usind CMB data to constraint primordial helicity. 

(d) Cosmological helicity 
 induces  parity-odd cross-correlations of the CMB fluctuations, 
which vanish for the case of a magnetic field or turbulent motions 
  without helicity 
(Lue et al. 1999, Pogosian et al. 2002, Caprini et al. 2004, 
Kahniashvili and Ratra 2005, Kahniashvili et al. 2005).
\footnote{This is not
  true for an homogeneous
magnetic field (Scoccola, Harari, and Mollerach 2004), or in the case 
of cosmological defects (Lepora 1998). }

\section{Polarized gravitational waves background}
The energy-momentum tensors corresponding to the magnetic field and turbulent motions have 
the anisotropic stress part which plays a source term role 
for graviational waves. If the parity violation (helicity) is present - 
the induced gravitational waves have a parity-odd spectrum (Lue et al. 1999),
 i.e. the gravitational waves background is circularly polarized  
(Caprini et al. 2004, Kahniashvili et al. 2005). 
Polarized gravitational waves also can be induced from quantum fluctuations 
through Chern-Simons coupling (Lyth, Quimbay and Rodriguez, 2005). 
The polarization degree of such a gravitational wave 
background strongly depends on the ratio between the helical and symmetric parts of the source two-point correlations function. 
We define the polarization degree as (Kahniashvili et al. 2005),
   \begin{eqnarray}
&&{\mathcal P}(k,t) =\frac{{\mathcal H}(k, t)}{H(k, t)}= 
\nonumber\\&&
 ~~~~=\frac {\langle h^{+ \star}({\mathbf k}, t)
h^{+}({\mathbf k'}, t) -
 h^{- \star}({\mathbf k},t) h^{-}({\mathbf k'},t) \rangle}
{\langle h^{+ \star}({\mathbf k}, t) h^{+}({\mathbf k'}, t) +
 h^{- \star}({\mathbf k},t) h^{-}({\mathbf k'},t) \rangle}
\label{degree}
\end{eqnarray}
Here $h^+$ and $h^-$ defines two states of the gravitational wave, '
  (right- and left-handed
 circularly  polarized gravitational waves),  $h_{ij}=h^+e^+_{ij} + h^-
e^-_{ij}$, where  $e^{\pm}_{ij}$ is polarization basis. 
$H ({k}, t)$ and ${\mathcal H}({k}, t)$ characterize
the gravitational wave  amplitude and polarization. 
 An axisymmetric stochastic vector source (non-helical turbulent
motion or any other non-helical vector field) induces unpolarized
GWs with $ |h^+({\bf k}, t)|  = |h^-({\bf k}, t)|$ 
(Deriagin et al. 1987; Durrer et al. 2000; 
Kosowsky et al. 2002; Dolgov, Grasso and Nicolis  et al. 2002; Lewis 2004; 
Caprini \& Durrer 2006).

For simplicity I present here the polarization degree of gravitational waves 
in the case of an helical turbulence model in which the turbulent 
motions ${\bf u}({\bf x}, t)$ 
are described by a time-dependent two-point correlation function, 
(Kosowsky, et al. 2002; Kahniashvili et al. 2005){\footnote{A 
 similar analysis for 
the polarization degree of gravitational waves induced by a stochastic 
 helical magnetic field is given  in (Caprini et al. 2004).}}, 
\begin{eqnarray}
&&\langle u^\star_i({\mathbf k}, t) u_j({\mathbf k'}, t')\rangle
=
={(2\pi)^3} \delta^{(3)}({\mathbf k}-{\mathbf k'})
\nonumber\\&&
~~~~~~~~~~~[P_{ij} F_S(k,
t-t^\prime)
+ i
\epsilon_{ijl} \hat{k}_l F_H(k, t-t^\prime)], \label{spectrumtime}
\end{eqnarray}
where  the time $t-t^\prime $
 dependence of the  functions $F_S$ and $F_H$
 reflects the assumption of time translation invariance. 
According our assumption energy is injected continuously, 
at  $t=t^\prime\in (t_{\rm{in}},
t_{\rm{fi}})$,  $F_{S}(k,0)=P_{S}(k)$ and $ F_{H}(k,0)=P_{H}(k)$.
 $P_S(k)$ and $P_H(k)$ are
the symmetric (related to the
kinetic energy density per unit enthalpy of the fluid)
 and helical (related to the  average
kinetic helicity $\langle \mathbf{u}
 \cdot (\nabla \times \mathbf{u}) \rangle$) parts of the
 velocity
power spectrum (Pogosian et al. 2002, Kahniashvili \& Ratra 2005). 
 We model the decay of helical turbulence 
by a monotonically decreasing functions $D_1(k)$ and $D_2(k)$, so 
 $F_S(k,t) = P_S(k) D_1(t)$ and $F_{H}(k,t)=P_{H}(k) D_{2}(t)$. 
We model the power spectra by  power laws,
$P_{S}(k) \propto k^{n_{S}}$ and $P_{H}(k) \propto k^{n_{H}}$. For
 non-helical hydrodynamical turbulence  the Kolmogorov spectrum
 has $n_S=-11/3$. The presence of hydrodynamical helicity makes 
 the situation more
 complex. Two possibilities have been discussed. First,
with a
forward cascade (from large  to small scales)
 of both energy and helicity
 (dominated by energy dissipation  on small scales) one has spectral
indices $n_S=-11/3$ and $n_H=-14/3$ (the helical Kolmogorov (HK)
spectrum), p.~243 of Lesieur, 1997.  Second, if helicity
 transfer and small-scale helicity dissipation dominate,
 $n_S=n_H=-13/3$ (the helicity transfer (HT) spectrum),  Moiseev \&  
Chkhetiani, 1996. Based on our assumptions (for 
the details see Kahniashvili et al. 2005)  
 we model the primordial  spectra as
$P_S(k)=S_0 k^{n_S}$ and $P_H(k)= A_0
k_S^{n_S-n_H}k^{n_H}$, where: (i) for the HK case $S_0=\pi^2C_k
{\bar\varepsilon }^{2/3}$ and $A_0=\pi^2C_k {\bar\delta}
/({\bar\varepsilon }^{1/3}k_S)$ (Ditlevsen \& Giuliani 2001), 
 implying $A_0/S_0 =
{\bar \delta}/({\bar \varepsilon}k_S)$; and,   (ii) for the HT case  $S_0 =
\pi^2 C_s {\bar\delta }^{2/3}$ and $A_0 = \pi^2 C_a {\bar\delta
}^{2/3}$ (Moissev \& Chkhetiani 1996).  Here
 ${\bar\varepsilon}$ and  ${\bar\delta}$ are the energy and mean helicity
 dissipation rates per unit enthalpy, and $C_k$ (the Kolmogorov constant),
 $C_s$, and $C_a$ are  constants of order unity. 
\begin{figure}[]
\mbox{\epsfig{figure=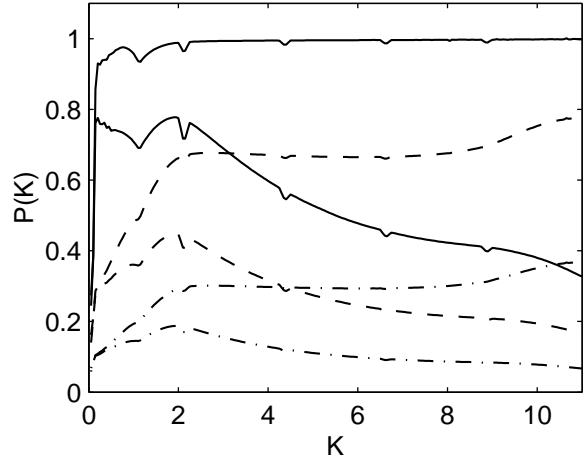,width=\linewidth,clip=}}
\caption{GW polarization degree ${\mathcal P}(K, t_{\rm{fi}})$,
Eq.~(\ref{degree}), as a function of scaled wave-number $K=k/k_S$
relative to the large-scale wave-number $k_S$ on which energy is
pumped into the turbulence. This is evaluated at time
$t_{\rm{fi}}$, after the turbulence has switched off, and remains
unchanged to the present epoch. It has been computed for a damping
wave-number $k_D=10k_S$. Three pairs of curves are shown. Solid
lines correspond to the amplitude ratio $A_0/S_0=1$, dashed lines
to $A_0/S_0=0.5$, and dot-dashed lines are for $A_0/S_0 =0.2$. The
upper line in each pair corresponds to HT turbulence with
$n_S=n_H=-13/3$ and the lower line to HK turbulence with
$n_S=-11/3$ and $n_H=-14/3$. Even for helical turbulence with
$A_0/S_0 \leq 0.5$,  for large wave-numbers $k \sim k_D$,
$n_S=n_H=-13/3$ is unlikely so the large $K$ part of the lower
dashed and dot-dashed HT curves are unrealistic. The large $k \sim
k_D$ decay of the HK curves is a consequence of vanishing helicity
transfer at large $k$. } 
\label{fig1}
\end{figure}
Figure 1 and other  numerical results show that for  maximal
helicity turbulence (when $A_0=S_0$) with equal spectral indices
$n_H = n_S<-3$, the polarization degree
${\mathcal P}(k) \simeq 1$ (upper solid line).
 For weaker helical turbulence (when $ A_0 <
S_0$) with $n_H \simeq  n_S<-3$,  ${\mathcal P(k)}
\rightarrow CA_0/S_0$, where $1<C(n_S,n_H)<2$ is a numerical factor that
depends on the spectral indices. For HT turbulence with
$n_S=n_H=-13/3$,  $C \approx 1.50$, while  for Iroshnikov-Kraichnan
MHD turbulence ($n_S=n_H=-7/2$), $C\approx1.39$. 
It is unlikely that such kind of polarized 
gravitational waves  will be detected in the near future, however, 
gravitational waves generated by helical turbulence will have an 
enough high degree of circular polarization and future detector 
configurations  may well be able to. On the other hand, the polarized 
gravitational waves might leave an observable traces on CMB aniotropies, in particular parity-violating cross crrelations between $B$-polarization and 
temperature and $E$-$B$ polarization.   

\section{CMB fluctuations}

Lets consider another parity violating source 
which might be present in the early universe --- a stochastic 
helical cosmological magnetoc field. 

Neglecting fluid back-reaction onto the magnetic field,
the spatial and temporal dependence of the
field separates, ${\mathbf B}(t,{\mathbf x})={\mathbf B}
({\mathbf x})/a^2$; here $a$ is the
 cosmological scale factor.  Assuming that the primordial plasma 
is a perfect conductor we model magnetic field  damping by an 
 ultraviolet cut-off wavenumber
$k_D=2\pi/\lambda_D$ (Subramanian and Barrow 1998). 
Gaussianly distributed an helical magnetic field two-point correlation function is 
\begin{eqnarray}
&&\langle B^\star_i({\mathbf k})B_j({\mathbf k'})\rangle
=\nonumber\\
&&=(2\pi)^3 \delta^{(3)}
({\mathbf k}-{\mathbf k'}) [P_{ij}({\mathbf{\hat k}}) P_B(k)  +
i \epsilon_{ijl} \hat{k}_l P_H(k)].
\label{spectrum}
\end{eqnarray}
where $P_B(k)$ and $P_H(k)$ are the symmetric and
helical parts of the magnetic field power
spectrum, assumed to be simple power laws on large scales \begin{eqnarray}
P_B(k) &\equiv & P_{B0}k^{n_B}=
\frac{2\pi^2 \lambda^3 B^2_\lambda}{\Gamma(n_B/2+3/2)}
(\lambda k)^{n_B},\nonumber\\
P_H(k) &\equiv &P_{H0}k^{n_H}=
\frac{ 2\pi^2 \lambda^3 H^2_\lambda}{\Gamma(n_H/2+2)}
(\lambda k)^{n_H},
\label{energy-spectrum-H}
\end{eqnarray} and 
vanishing on small scales when $k>k_D$. 
$B_\lambda^2$ is the squared smoothed magnetic field amplitude at the 
$\lambda$ scale. 

The
symmetric part of the magnetic field spectrum can be
reconstructed from  measurements of Faraday 
 rotation of the CMB polarization
 plane 
(Kosowsky et al. 2005). This is because
magnetic helicity does not contribute to the Faraday rotation
effect (Ensslin \& Vogt 2003, Campanelli et al. 2004, 
Kosowsky et al. 2005 ). 
Faraday rotation by an helical magnetic field induces 
a $B$-polarization signal that peaks at very high multipole 
number, $l \sim 15 000$  (see Fig. 2). 
The position of this peak 
makes it  possible to distinguish the Faraday-rotation 
 induced $B$-polarization 
 signal from the signal arising from the presence of the vector and 
 tensor modes which  peak around $l \sim 2000$ (Lewis 2004; Challinor \& 
Lewis 2005). 
\begin{figure}[]
\mbox{\epsfig{figure=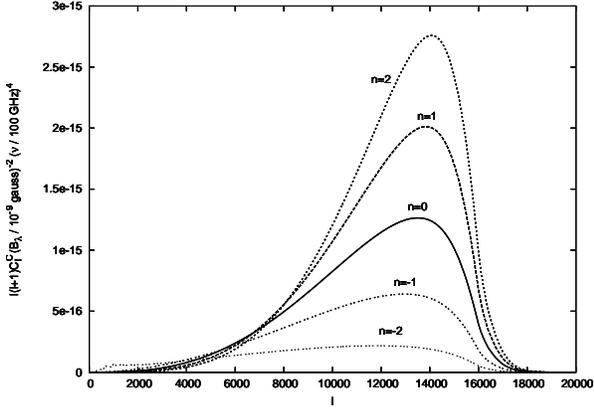,angle=270,width=\linewidth,clip=}}
\caption{The $B$-polarization power spectrum of the microwave background
induced by the Faraday rotation field. The curves in
order of decreasing amplitude on the right side of the plot correspond
to magnetic field power spectral indices $n_B=2$, 1, 0, $-1$, and $-2$.
The magnetic fields have been normalized to a nanogauss at the smoothing
scale $\lambda = 1$ Mpc.}
\label{fig2}
\end{figure}

The helical part of the magnetic field
spectrum induces parity-odd cross correlations between
temperature and $B$-polarization anisotropies,
 and between $E$- and $B$-polarization anisotropies. 
Below I discuss explicity these parity-odd cross correlations. 

For our computations we use the formalism  by Mack et al. (2002),
extending it to account for magnetic field helicity. To compute
CMB temperature and polarization anisotropy power spectra we use
the total angular momentum method by Hu and White (1997). Our
results (Caprini et al. 2004, Kahniashvili \& Ratra 2006) 
 are obtained using analytic approximations and for the vector mode 
 are valid
for $l<500$. We emphasis that for the tensor mode the usage of our 
approximations are limited by $l<60$ because of the fast decay of 
the gravitational wave source in the matter-dominated epoch. 
Our results might be presented in terms of a ratio between CMB
fluctuation contributions from  the symmetric and   helical parts
of the magnetic field power spectrum.

To obtain the magnetic field source  terms in the equations for vector
(transverse
peculiar velocity)  and tensor (gravitational waves) metric perturbations
  we need to
extract the transverse vector and tensor parts of the magnetic
field stress-energy tensor $\tau_{ij}({\mathbf{k}})$. This is done
through $ \Pi_{ij}^{(V)}({\mathbf{k}})=(P_{ib}({\mathbf{\hat k}})
\hat{k}_j+P_{jb}({\mathbf{\hat k}})\hat{k}_i)\hat{k}_a
\tau_{ab}({\mathbf{k}}) $ (for vector perturbations) and  $
\Pi_{ij}^{(T)} ({\mathbf {k}})=[ P_{ia}({\mathbf{\hat
k}})P_{jb}({\mathbf{\hat k}}) -\frac{1}{2} P_{ij}({\mathbf{\hat
k}}) P_{ab}({\mathbf{\hat k}}) ]\tau_{ab}({\mathbf{ k}})$ (for
tensor perturbations); for details see Mack et al. (2002).
In both cases (vector and tensor perturbations)
the contribution of magnetic field
helicity to
the symmetric part of the magnetic source is negative.

As we have already noted, a possible way to detect 
magnetic helicity directly from CMB fluctuation data
 is  to observe the  
parity-odd CMB fluctuation cross-correlations. As an additional cross-check 
 it may be possible to detect    
the effects that magnetic  helicity has on
 parity-even CMB fluctuations (Caprini et al. 2004, 
 Kahniashvili and Ratra 2005). Since we find (Kahniashvili \& Ratra 2006) 
that magnetic field induced 
 density perturbations are independent on  magnetic helicity,   
if one can extract the scalar mode contribution from the total 
magnetic field sourced CMB fluctuations, that will allow 
for a determination of 
the symmetric part of the magnetic field spectrum 
(see also Giovannini 2006a,b),  and result in a more accurate 
estimate of magnetic helicity from parity-odd CMB anisotropies.

At large angular scales ($l<100$) where the contribution from
the tensor mode is significant, for $n_B+n_H>-2$ the  vector mode
${C_l^{\Theta B (V)}}$ and the tensor mode
${C_l^{\Theta B (T)}}$
have the same $l$ dependence $\propto l^2$.
For all other values of
spectral indexes $n_{B}$ and $n_{H}$, the growth rate (with $l$) of
 ${C_l^{\Theta B (V)}}$  is faster than
${C_l^{\Theta B (T)}}$. The ratio between  temperature--$B$-polarization
signals from vector and tensor modes is independent
of the amplitudes
of the average magnetic field ($B_\lambda$) and
average magnetic helicity ($H_\lambda$).

For small angular scales ($l> 100$) where the tensor mode signal vanishes, for
a maximally helical magnetic field with $n_H \simeq n_B$, due
to the suppression factor $L_{\gamma,{\rm{dec}}}/\eta_0$ (where 
$L_{\gamma,{\rm{dec}}}$ is the photon mean free path  at decoupling and  
 $\eta_0$ conformal time today)
the temperature-$E$-polarization cross-correlation power spectrum, 
$C_l^{\Theta E}$, is smaller than the  temperature-$B$-polarization
cross-correlation power spectrum, 
$C_l^{\Theta B}$, but both are $ \propto l^2$, if $n_B+n_H>-5$. The same
suppression factor makes $C_l^{\Theta B}$   smaller than
$C_l^{\Theta \Theta}$. For an arbitrary helical field  
 $C_l^{\Theta B}/C_l^{\Theta E}$  depends on the ratio $(P_{H0}/
P_{B0})k_D^{n_H-n_B}$ and order unity  prefactors that depend on
$n_B$ and  $n_H$.
 A dependence on $l$ appears
only if $n_B+n_H<-5$, when  the ratio,
$C_l^{\Theta B}/C_l^{\Theta E}$ decreases as  $\propto l^{n_B+n_H+5}$
(Kahniashvili \& Ratra 2005).

For a tensor mode signal at large angular scales ($l<100$),
 $E$- and $B$-polarization cross-correlation $C_l^{EB}$ 
is of the same order of magnitude as the tensor
mode temperature--$B$-polarization anisotropy cross-correlation
 spectrum, $C_l^{\Theta B}$ (Caprini et al. 2004).
 The situation is different for a vector mode which survives up to small
angular scales (e.g.,
Subramanian \& Barrow 1998, Mack et al. 2002, Lewis 2004, Giovannini 2006a).
  In this case,
the $E$- and $B$-polarization anisotropy cross-correlation power
spectrum has a suppression factor of $kL_{\gamma, \rm{dec}}$
implying that $C_l^{EB} \ll C_l^{\Theta B}$. This is
consistent   with the result of Hu and White (1997).

\section{Conclusion} 
I have discussed  the  cosmological effects of primordial  helicity.  
 In particular, I examined 
 CMB parity-violating fluctuations that arise  
 from helical sources. These CMB fluctuations  should be 
detectable (if the current magnetic
 field amplitude is at least
$10^{-10}$ or $10^{-9}$ G on Mpc scales --- 
 such a magnetic field can be generated during inflation from quantum fluctuations, see Ratra 1992, 
Bamba \& Yokoyama 2004) 
 by near future CMB polarization measurements (from WMAP, 
 PLANK, CMBPol and others). 
 As a specific imprint of primordial kinetic helicity I discussed  
polarization of relic gravitational waves, possibly detectable by future 
space missions (Smith, Kamionkowski and Cooray 2006, 
Chongchitnan \& Efstatiou 2006).

The author thanks the organizers of the cosmology workshop at Irvine 
for hospitality, and acknowledges her collaborators C. Caprini, R. Durrer,
G. Gogoberidze, A. Kosowsky, G. Lavrelashvili, A. Mack, and B. Ratra. 
The author also thanks A.~Brandenburg, A. Dolgov, M. Giovannini, D. Grasso,
K. Jedamzik, and T. Vachaspati for discussions.
 This work is supported by DOE
EPSCoR grant DE-FG02-00ER45824.


\end{document}